\documentclass[twocolumn]{aastex701}

\usepackage{graphicx}
\usepackage{amsmath}
\usepackage{hyperref}
\usepackage{longtable}
\usepackage{booktabs}

\newcommand{\te}{t_{\rm E}}
\newcommand{\thetae}{\theta_{\rm E}}

\newcommand{\pie}{\pi_{\rm E}}

\newcommand{\dl}{D_{\rm L}}
\newcommand{\ds}{D_{\rm S}}
\def\e{{\rm E}}

\definecolor{brown}{rgb}{0.59, 0.29, 0.0}
\definecolor{darkgreen}{rgb}{0.0, 0.42, 0.24}
\definecolor{darkblue}{rgb}{0.01, 0.31, 0.59}
\definecolor{blue}{rgb}{0.0,0.0,1.0}
\definecolor{green}{rgb}{0.0,1.0,0.0}

\begin{document}

\title{Four Cold Giant Planets Discovered by High-Cadence Microlensing Surveys}
\shorttitle{Four Cold Giant Microlensing Planets}

\author{Cheongho Han}
\affiliation{Department of Physics, Chungbuk National University, Cheongju 28644, Republic of Korea}
\email{cheongho@astroph.chungbuk.ac.kr}
\author{Chung-Uk Lee}
\affiliation{Korea Astronomy and Space Science Institute, Daejon 34055, Republic of Korea}
\email{leecu@kasi.re.kr}
\author{Andrzej Udalski} 
\affiliation{Astronomical Observatory, University of Warsaw, Al.~Ujazdowskie 4, 00-478 Warszawa, Poland}
\email{udalski@astrouw.edu.pl} 
\author{Ian A. Bond}
\affiliation{School of Mathematical and Computational Sciences, Massey University, Auckland 0745, New Zealand}
\email{i.a.bond@massey.ac.nz}
\collaboration{14}{(Leading authors)}
\author{Michael D. Albrow}   
\affiliation{University of Canterbury, Department of Physics and Astronomy, Private Bag 4800, Christchurch 8020, New Zealand}
\email{michael.albrow@canterbury.ac.nz}
\author{Sun-Ju Chung}
\affiliation{Korea Astronomy and Space Science Institute, Daejon 34055, Republic of Korea}
\email{sjchung@kasi.re.kr}
\author{Andrew Gould}
\affiliation{Department of Astronomy, Ohio State University, 140 West 18th Ave., Columbus, OH 43210, USA}
\email{gould.34@osu.edu}
\author{Youn Kil Jung}
\affiliation{Korea Astronomy and Space Science Institute, Daejon 34055, Republic of Korea}
\affiliation{University of Science and Technology, Daejeon 34113, Republic of Korea}
\email{younkil21@gmail.com}
\author{Kyu-Ha~Hwang}
\affiliation{Korea Astronomy and Space Science Institute, Daejon 34055, Republic of Korea}
\email{kyuha@kasi.re.kr}
\author{Yoon-Hyun Ryu}
\affiliation{Korea Astronomy and Space Science Institute, Daejon 34055, Republic of Korea}
\email{yhryu@kasi.re.kr}
\author{Yossi Shvartzvald}
\affiliation{Department of Particle Physics and Astrophysics, Weizmann Institute of Science, Rehovot 76100, Israel}
\email{yossishv@gmail.com}
\author{In-Gu Shin}
\affiliation{Department of Astronomy, Westlake University, Hangzhou 310030, Zhejiang Province, China}
\email{ingushin@gmail.com}
\author{Jennifer C. Yee}
\affiliation{Center for Astrophysics $|$ Harvard \& Smithsonian 60 Garden St., Cambridge, MA 02138, USA}
\email{jyee@cfa.harvard.edu}
\author{Weicheng Zang}
\affiliation{Department of Astronomy, Westlake University, Hangzhou 310030, Zhejiang Province, China}
\email{zangweicheng@westlake.edu.cn}
\author{Hongjing Yang}
\affiliation{Department of Astronomy, Westlake University, Hangzhou 310030, Zhejiang Province, China}
\email{yanghongjing@westlake.edu.cn}
\author{Doeon Kim}
\affiliation{Department of Physics, Chungbuk National University, Cheongju 28644, Republic of Korea}
\email{qso21@hanmail.net}
\author{Dong-Jin Kim}
\affiliation{Korea Astronomy and Space Science Institute, Daejon 34055, Republic of Korea}
\email{keaton03@kasi.re.kr}
\author{Seung-Lee Kim}
\affiliation{Korea Astronomy and Space Science Institute, Daejon 34055, Republic of Korea}
\email{slkim@kasi.re.kr}
\author{Dong-Joo Lee}
\affiliation{Korea Astronomy and Space Science Institute, Daejon 34055, Republic of Korea}
\email{marin678@kasi.re.kr}
\author{Sang-Mok Cha}
\affiliation{Korea Astronomy and Space Science Institute, Daejon 34055, Republic of Korea}
\email{chasm@kasi.re.kr}
\author{Yongseok Lee}
\affiliation{Korea Astronomy and Space Science Institute, Daejon 34055, Republic of Korea}
\email{yslee@kasi.re.kr}
\author{Byeong-Gon Park}
\affiliation{Korea Astronomy and Space Science Institute, Daejon 34055, Republic of Korea}
\email{bgpark@kasi.re.kr}
\author{Richard W. Pogge}
\affiliation{Department of Astronomy, Ohio State University, 140 West 18th Ave., Columbus, OH 43210, USA}
\email{pogge.1@osu.edu}
\collaboration{20}{(KMTNet Collaboration)}
\author{Przemek Mr{\'o}z}
\affiliation{Astronomical Observatory, University of Warsaw, Al.~Ujazdowskie 4, 00-478 Warszawa, Poland}
\email{pmroz@astrouw.edu.pl}
\author{Micha{\l} K. Szyma{\'n}ski}
\affiliation{Astronomical Observatory, University of Warsaw, Al.~Ujazdowskie 4, 00-478 Warszawa, Poland}
\email{msz@astrouw.edu.pl}
\author{Jan Skowron}
\affiliation{Astronomical Observatory, University of Warsaw, Al.~Ujazdowskie 4, 00-478 Warszawa, Poland}
\email{jskowron@astrouw.edu.pl}
\author{Rados{\l}aw Poleski} 
\affiliation{Astronomical Observatory, University of Warsaw, Al.~Ujazdowskie 4, 00-478 Warszawa, Poland}
\email{radek.poleski@gmail.co}
\author{Igor Soszy{\'n}ski}
\affiliation{Astronomical Observatory, University of Warsaw, Al.~Ujazdowskie 4, 00-478 Warszawa, Poland}
\email{soszynsk@astrouw.edu.pl}
\author{Pawe{\l} Pietrukowicz}
\affiliation{Astronomical Observatory, University of Warsaw, Al.~Ujazdowskie 4, 00-478 Warszawa, Poland}
\email{pietruk@astrouw.edu.pl}
\author{Szymon Koz{\l}owski} 
\affiliation{Astronomical Observatory, University of Warsaw, Al.~Ujazdowskie 4, 00-478 Warszawa, Poland}
\email{simkoz@astrouw.edu.pl}
\author{Krzysztof A. Rybicki}
\affiliation{Astronomical Observatory, University of Warsaw, Al.~Ujazdowskie 4, 00-478 Warszawa, Poland}
\email{krybicki@astrouw.edu.pl}
\author{Patryk Iwanek}
\affiliation{Astronomical Observatory, University of Warsaw, Al.~Ujazdowskie 4, 00-478 Warszawa, Poland}
\email{piwanek@astrouw.edu.pl}
\author{Krzysztof Ulaczyk}
\affiliation{Department of Physics, University of Warwick, Gibbet Hill Road, Coventry, CV4 7AL, UK}
\email{kulaczyk@astrouw.edu.pl}
\author{Marcin Wrona}
\affiliation{Astronomical Observatory, University of Warsaw, Al.~Ujazdowskie 4, 00-478 Warszawa, Poland}
\affiliation{Villanova University, Department of Astrophysics and Planetary Sciences, 800 Lancaster Ave., Villanova, PA 19085, USA}
\email{mwrona@astrouw.edu.pl}
\author{Mariusz Gromadzki}          
\affiliation{Astronomical Observatory, University of Warsaw, Al.~Ujazdowskie 4, 00-478 Warszawa, Poland}
\email{marg@astrouw.edu.pl}
\author{Mateusz J. Mr{\'o}z} 
\affiliation{Astronomical Observatory, University of Warsaw, Al.~Ujazdowskie 4, 00-478 Warszawa, Poland}
\email{mmroz@astrouw.edu.pl}
\collaboration{100}{(The OGLE Collaboration)}
\author{David P.~Bennett}
\affiliation{Code 667, NASA Goddard Space Flight Center, Greenbelt, MD 20771, USA}
\affiliation{Department of Astronomy, University of Maryland, College Park, MD 20742, USA}
\email{bennett.moa@gmail.com}
\author{Chihiro Ueda}
\affiliation{Department of Earth and Space Science, Graduate School of Science, The University of Osaka, Toyonaka, Osaka 560-0043, Japan}
\email{Ueda@iral.ess.sci.osaka-u.ac.jp}
\author{Ryusei Hamada}
\affiliation{Department of Earth and Space Science, Graduate School of Science, The University of Osaka, Toyonaka, Osaka 560-0043, Japan}
\email{hryusei@iral.ess.sci.osaka-u.ac.jp}
\author{Yuki Hirao}
\affiliation{Institute of Astronomy, Graduate School of Science, The University of Tokyo, 2-21-1 Osawa, Mitaka, Tokyo 181-0015, Japan}
\email{hirao@ioa.s.u-tokyo.ac.jp}
\author{Shuma Makida}
\affiliation{Department of Earth and Space Science, Graduate School of Science, The University of Osaka, Toyonaka, Osaka 560-0043, Japan}
\email{makida@iral.ess.sci.osaka-u.ac.jp}
\author{Shota Miyazaki}
\affiliation{Institute of Space and Astronautical Science, Japan Aerospace Exploration Agency, 3-1-1 Yoshinodai, Chuo, Sagamihara, Kanagawa 252-5210, Japan}
\email{miyazaki@ir.isas.jaxa.jp}
\author{Tutumi Nagai}
\affiliation{Department of Earth and Space Science, Graduate School of Science, The University of Osaka, Toyonaka, Osaka 560-0043, Japan}
\email{nagai@iral.ess.sci.osaka-u.ac.jp}
\author{Seiya Nakayama}
\affiliation{Department of Earth and Space Science, Graduate School of Science, The University of Osaka, Toyonaka, Osaka 560-0043, Japan}
\email{nakayama@iral.ess.sci.osaka-u.ac.jp}
\author{Kansuke Nunota}
\affiliation{Department of Earth and Space Science, Graduate School of Science, The University of Osaka, Toyonaka, Osaka 560-0043, Japan}
\email{nunota@iral.ess.sci.osaka-u.ac.jp}
\author{Ryo Ogawa}
\affiliation{Department of Earth and Space Science, Graduate School of Science, The University of Osaka, Toyonaka, Osaka 560-0043, Japan}
\email{rogawa@iral.ess.sci.osaka-u.ac.jp}
\author{Ryunosuke Oishi}
\affiliation{Department of Earth and Space Science, Graduate School of Science, The University of Osaka, Toyonaka, Osaka 560-0043, Japan}
\email{oishi@iral.ess.sci.osaka-u.ac.jp}
\author{Hideaki Ose}
\affiliation{Department of Earth and Space Science, Graduate School of Science, The University of Osaka, Toyonaka, Osaka 560-0043, Japan}
\email{ose@iral.ess.sci.osaka-u.ac.jp}
\author[0000-0001-5069-319X]{Nicholas J. Rattenbury}
\affiliation{Department of Physics, University of Auckland, Private Bag 92019, Auckland, New Zealand}
\email{n.rattenbury@auckland.ac.nz}
\author[0000-0002-1228-4122]{Yuki K. Satoh}
\affiliation{College of Science and Engineering, Kanto Gakuin University, Yokohama, Kanagawa-2-236-8501, Japan}
\email{yukisato@kanto-gakuin.ac.jp}
\author{Takahiro Sumi}
\affiliation{Department of Earth and Space Science, Graduate School of Science, The University of Osaka, Toyonaka, Osaka 560-0043, Japan}
\email{sumi@ess.sci.osaka-u.ac.jp}
\author{Daisuke Suzuki}
\affiliation{Department of Earth and Space Science, Graduate School of Science, The University of Osaka, Toyonaka, Osaka 560-0043, Japan}
\email{dsuzuki@ir.isas.jaxa.jp}
\author[0000-0002-6510-0681]{Motohide Tamura}
\affiliation{Astrobiology Center, 2-21-1 Osawa, Mitaka-shi, Tokyo 181-8588, Japan}
\affiliation{Department of Astronomy, University of Tokyo, 7-3-1 Hongo, Bunkyo-ku, Tokyo 113-0033, Japan}
\email{motohide.tamura@astron.s.u-tokyo.ac.jp}
\author{Takuto Tamaoki}
\affiliation{Department of Earth and Space Science, Graduate School of Science, The University of Osaka, Toyonaka, Osaka 560-0043, Japan}
\email{tamaoki@iral.ess.sci.osaka-u.ac.jp}
\author{Hibiki Yama}
\affiliation{Department of Earth and Space Science, Graduate School of Science, The University of Osaka, Toyonaka, Osaka 560-0043, Japan}
\email{yama@iral.ess.sci.osaka-u.ac.jp}
\collaboration{100}{(The PRIME Collaboration)}
\correspondingauthor{\texttt{cheongho@astroph.chungbuk.ac.kr}}
\correspondingauthor{\texttt{leecu@kasi.re.kr}}

\begin{abstract}
We report the discovery of four cold giant planets identified through the analysis of microlensing 
events detected by high-cadence surveys: OGLE-2016-BLG-0261, KMT-2025-BLG-0026, KMT-2025-BLG-0030, 
and KMT-2025-BLG-2272. The planetary signals appear as short-duration anomalies in the light curves 
and are well described by binary-lens single-source models with mass ratios between the lens components 
of order $q \sim 10^{-3}$. Finite-source effects are securely measured in three out of four events, 
enabling determinations of the angular Einstein radius. A Bayesian analysis incorporating the measured 
event timescale and angular Einstein radius yields host masses of $\sim 0.07$--$0.6~M_\odot$ and 
companion masses of $\sim 0.2$--$2.5~M_{\rm J}$, confirming that all companions lie in the giant-planet 
regime.  The projected separations are $\sim 0.7$--$6$ au, placing all planets at or beyond the snow 
lines of their host stars.  The inferred lens distances span $\sim 6.6$–$7.9$ kpc, with all systems 
consistent with bulge lenses.  These detections expand the sample of cold giant planets from homogeneous 
high-cadence surveys and highlight the sensitivity of microlensing to planetary systems beyond the snow 
line, providing further constraints on the occurrence and properties of giant planets around low-mass 
stars.  
\end{abstract}


\section{Introduction \label{sec:one}} 

Gravitational microlensing provides a unique probe of exoplanets located at separations of order 
a few astronomical units, corresponding to regions beyond the snow line where giant planet
formation is expected to be most efficient. In contrast to the radial velocity and transit methods,
which are primarily sensitive to close-in planets, microlensing surveys enable a direct census of
cold planetary systems, including those orbiting low-mass and distant hosts in both the Galactic
disk and bulge. As a result, microlensing plays a critical role in establishing a complete picture 
of planet populations across a wide range of orbital separations and host environments
\citep{Gaudi2012}.

Recent advances in wide-field, high-cadence surveys have significantly improved the sensitivity 
of microlensing experiments to short-duration, low-amplitude planetary signals \citep{Bond2002, 
Udalski2015, Kim2016}. In particular, continuous monitoring with cadences down to several tens 
of minutes has enabled the detection of subtle perturbations that would have been missed in earlier 
survey-plus-follow-up strategies \citep{Griest1998, Gould2010}. These developments have led to a 
steady increase in the number of detected planets and, importantly, to the construction of more 
homogeneous samples better suited for statistical analyses of planet occurrence \citep{Suzuki2016, 
Zang2025}.

\begin{deluxetable}{lllllll}
\tabletypesize{\footnotesize}
\tablewidth{0pt}
\tablecaption{Coordinates and event ID correspondence. \label{table:one}}
\tablehead{
\multicolumn{1}{c}{KMTNet ID}                        &
\multicolumn{1}{c}{$(\alpha, \delta)_{\rm J2000}$}   &
\multicolumn{1}{c}{$(l, b)$}                         &
\multicolumn{1}{c}{OGLE ID }                         &        
\multicolumn{1}{c}{PRIME ID}                       
}
\startdata
     KMT-2016-BLG-1679   &   (17:51:15.89, -30:24:30.60)   &  ($-0\fdg6290$, $-1\fdg8098)$  &  {\bf OGLE-2016-BLG-0261}   &   \nodata             \\
{\bf KMT-2025-BLG-0026}  &   (17:27:03.96, -29:16:26.69)   &  ($-2\fdg4666$, $+3\fdg2341)$  &  OGLE-2025-BLG-0156         &   \nodata             \\
{\bf KMT-2025-BLG-0030}  &   (17:30:20.82, -29:28:54.23)   &  ($-2\fdg2447$, $+2\fdg5236)$  &  OGLE-2025-BLG-0153         &   PRIME-2025-BLG0041  \\
{\bf KMT-2025-BLG-2272}  &   (17:58:18.04, -27:46:25.79)   &  ($+2\fdg4226$, $-1\fdg8071)$  &  \nodata                    &   \nodata             
\enddata                                                                                                                            
\tablecomments{The event IDs shown in bold indicate the identifiers adopted for the events.}  
\end{deluxetable}                                                                                                          

Despite this progress, the population of giant planets beyond the snow line remains only partially
characterized. Of particular interest is the distribution of planet masses in the sub-Saturn to
super-Jupiter regime, where previous microlensing studies have suggested a break in the mass-ratio
function \citep{Suzuki2016, Zang2025}. In addition, the frequency of giant planets orbiting low-mass 
stars remains a key test of planet formation theories, as standard core accretion models predict 
reduced formation efficiency in such environments \citep{Ida2004}. The dependence of planet occurrence 
on Galactic location, reflecting differences in metallicity and stellar population between the disk 
and bulge, also remains an open question.

In this work, we present the discovery and analysis of four giant planets identified from high-cadence 
microlensing survey data. The planets span masses from sub-Saturn to super-Jupiter and orbit host 
stars ranging from very low-mass to near-solar mass. Their inferred projected separations place all 
systems beyond the snow line, making them well suited for probing the cold planet population. This 
sample adds to the growing set of uniformly detected microlensing planets and provides new constraints 
on the demographics of giant planets in the outer regions of planetary systems. In particular, it 
offers insight into the efficiency of giant planet formation around low-mass hosts and the Galactic 
distribution of planetary systems, helping to refine our understanding of planet formation and 
evolution in diverse environments.

The structure of this paper is as follows. In Sect.~\ref{sec:two}, we describe the observations 
and data reduction procedures for the four microlensing events analyzed in this work, including 
details of the survey coverage and photometric processing. In Sect.~\ref{sec:three}, we present 
the light-curve modeling, outline the adopted modeling strategy, and discuss the lensing solutions 
identified for each event, along with potential degeneracies. In Sect.~\ref{sec:four}, we determine 
the angular Einstein radii based on the characterization of the source stars and the measurement 
of finite-source effects. In Sect.~\ref{sec:five}, we estimate the physical parameters of the 
lens systems through a Bayesian analysis incorporating the measured observables. Finally, in 
Sect.~\ref{sec:six}, we summarize our main results and discuss their implications for the 
population and formation of cold giant planets.

\section{Observations and data} \label{sec:two}

The planets reported in this work were identified through the analyses of four microlensing 
events including KMT-2016-BLG-1679, KMT-2025-BLG-0026, KMT-2025-BLG-0030, and KMT-2025-BLG-2272. 
These events were detected via high-cadence photometric observations conducted by the microlensing 
survey collaborations of the Korea Microlensing Telescope Network 
\citep[KMTNet;][]{Kim2016}, the Optical Gravitational Lensing Experiment 
\citep[OGLE;][]{Udalski2015}, and the Prime-focus Infrared Microlensing Experiment 
\citep[PRIME;][]{Sumi2025}. The equatorial and Galactic coordinates for each event are provided 
in Table~\ref{table:one}.

Among these events, three were monitored by multiple survey groups. KMT-2016-BLG-1679 and 
KMT-2025-BLG-0026 were observed by both KMTNet and OGLE, while KMT-2025-BLG-0030 was observed 
by KMTNet, OGLE, and PRIME.  The remaining event, KMT-2025-BLG-2272, was observed exclusively 
by KMTNet.  For events detected by multiple surveys, we adopt the designation assigned by the 
first discovery group, as indicated in bold in Table~\ref{table:one}.

\begin{deluxetable}{lllllll}
\tablewidth{0pt}
\tablecaption{Error bar rescling parameters. \label{table:two}}
\tablehead{
\multicolumn{1}{c}{Data set}                 &
\multicolumn{1}{c}{$k$}                      &
\multicolumn{1}{c}{$\sigma_{\rm min}$ (mag)}                       
}
\startdata
 OGLE-2016-BLG-0261  &                    &                     \\
 \hskip4pt OGLE      & \ \ \ 1.517\ \ \   & \ \ \ 0.02\ \ \     \\    
 \hskip4pt KMTC01    & \ \ \ 1.003\ \ \   & \ \ \ 0.02\ \ \     \\    
 \hskip4pt KMTC41    & \ \ \ 1.002\ \ \   & \ \ \ 0.02\ \ \     \\    
 \hskip4pt KMTS01    & \ \ \ 1.034\ \ \   & \ \ \ 0.02\ \ \     \\    
 \hskip4pt KMTS41    & \ \ \ 0.938\ \ \   & \ \ \ 0.02\ \ \     \\    
 \hskip4pt KMTA01    & \ \ \ 0.943\ \ \   & \ \ \ 0.02\ \ \     \\    
 \hskip4pt KMTA41    & \ \ \ 0.894\ \ \   & \ \ \ 0.02\ \ \     \\    
\hline
 KMT-2025-BLG-0026   &                    &                     \\
 \hskip4pt OGLE      & \ \ \ 0.267\ \ \   & \ \ \ 0.02\ \ \     \\    
 \hskip4pt KMTC11    & \ \ \ 0.476\ \ \   & \ \ \ 0.03\ \ \     \\    
 \hskip4pt KMTS11    & \ \ \ 0.698\ \ \   & \ \ \ 0.01\ \ \     \\    
 \hskip4pt KMTA11    & \ \ \ 0.563\ \ \   & \ \ \ 0.02\ \ \     \\    
\hline
 KMT-2025-BLG-0030   &                    &                    \\
 \hskip4pt OGLE      & \ \ \ 1.002\ \ \   & \ \ \ 0.02\ \ \     \\    
 \hskip4pt KMTC11    & \ \ \ 0.529\ \ \   & \ \ \ 0.02\ \ \     \\    
 \hskip4pt KMTS11    & \ \ \ 0.600\ \ \   & \ \ \ 0.02\ \ \     \\    
 \hskip4pt KMTA11    & \ \ \ 0.702\ \ \   & \ \ \ 0.02\ \ \     \\    
 \hskip4pt PRIME     & \ \ \ 0.998\ \ \   & \ \ \ 0.02\ \ \     \\    
\hline
 KMT-2025-BLG-2272   &                    &                     \\
 \hskip4pt KMTC03    & \ \ \ 0.917\ \ \   & \ \ \ 0.02\ \ \     \\    
 \hskip4pt KMTC43    & \ \ \ 0.959\ \ \   & \ \ \ 0.02\ \ \     \\    
 \hskip4pt KMTS03    & \ \ \ 0.984\ \ \   & \ \ \ 0.02\ \ \     \\    
 \hskip4pt KMTS43    & \ \ \ 0.922\ \ \   & \ \ \ 0.02\ \ \     \\    
 \hskip4pt KMTA03    & \ \ \ 0.714\ \ \   & \ \ \ 0.02\ \ \     \\    
 \hskip4pt KMTA43    & \ \ \ 0.806\ \ \   & \ \ \ 0.02\ \ \     \\  
\enddata
\end{deluxetable}

Observations were carried out by globally distributed wide-field microlensing survey networks 
that provide complementary temporal coverage. KMTNet operates three identical 1.6~m telescopes
located at Cerro Tololo Inter-American Observatory in Chile (KMTC), the South African
Astronomical Observatory in South Africa (KMTS), and Siding Spring Observatory in Australia
(KMTA). This global network enables near-continuous monitoring of Galactic bulge fields with
cadences ranging from $\sim 0.25$ to $5$~hr, depending on the field.  The OGLE survey carries 
out optical observations using the 1.3~m Warsaw Telescope at Las Campanas Observatory in Chile, 
while the PRIME survey performs near-infrared (NIR) observations using a 1.8~m telescope at the 
South African Astronomical Observatory.  The camera fields of view are approximately $4.0~\rm deg^2$ 
for KMTNet, $1.4~\rm deg^2$ for OGLE, and $1.56~\rm deg^2$ for PRIME, enabling efficient wide-field 
monitoring of dense stellar fields toward the Galactic bulge. Observations from KMTNet and OGLE were 
primarily obtained in the standard Cousins $I$ band, with occasional $V$-band measurements used to 
determine source colors, while PRIME observations were conducted mainly in the $H$ band with occasional 
$J$-band data.

Photometric reductions for all datasets were carried out using the pipelines of the individual 
survey groups, each based on the Difference Image Analysis technique \citep{Alard1998, Tomaney1996}. 
The KMTNet data were processed with a pySIS-based pipeline developed by \citet{Albrow2009}. The 
OGLE and PRIME data were reduced using their respective pipelines: the OGLE reduction follows 
\citet{Wozniak2000}, while the PRIME pipeline adopts the methods described by \citet{Sumi2025}. 
These approaches are optimized for high-precision photometry in the crowded stellar fields of 
the Galactic bulge.

To ensure a consistent joint analysis, we renormalized the photometric uncertainties of each 
dataset following the standard procedure in which the error bars are rescaled as $\sigma^\prime 
= k(\sigma^2 + \sigma_{\rm min}^2)^{1/2}$ \citep{Yee2012}.  In this formulation, the parameter 
$\sigma_{\rm min}$ is determined for each dataset such that the cumulative $\chi^2$ distribution 
is approximately linear as a function of magnification, and the coefficient $k$ is a scaling 
factor chosen so that the reduced $\chi^2$ for each dataset is unity. This normalization ensures 
that all datasets are properly weighted in the modeling process.  Table~\ref{table:two} lists 
the error-bar rescaling parameters for the individual data sets of the analyzed events.

\section{Analysis} \label{sec:three}

All four events analyzed in this work exhibit short-duration anomalies superposed on otherwise 
smooth microlensing light curves. Such deviations are often indicative of low-mass companions 
to the lenses. We therefore investigate their origin by modeling the light curves with binary-lens 
single-source (2L1S) configurations.

A 2L1S model is described by seven principal lensing parameters. The first three, $(t_0, u_0,
\te)$, characterize the underlying single-lens event: $t_0$ is the time of the closest approach
between the source and the lens, $u_0$ is the lens-source separation at that moment normalized
to the angular Einstein radius ($\thetae$), and $\te$ is the Einstein timescale. The remaining 
four parameters, $(s, q, \alpha, \rho)$, describe the binary nature of the lens and the finite 
extent of the source: $s$ is the projected separation between the lens components normalized to 
$\thetae$, $q$ is the mass ratio between the companion and the primary lens, $\alpha$ is the 
angle between the source trajectory and the binary axis, measured clockwise from the binary axis, 
and $\rho$ is the angular source radius normalized to $\thetae$.  For events in which finite-source 
effects are important, $\rho$ can be robustly constrained from the anomaly morphology.

\begin{figure*}[t]
\centering
\includegraphics[width=13.0cm]{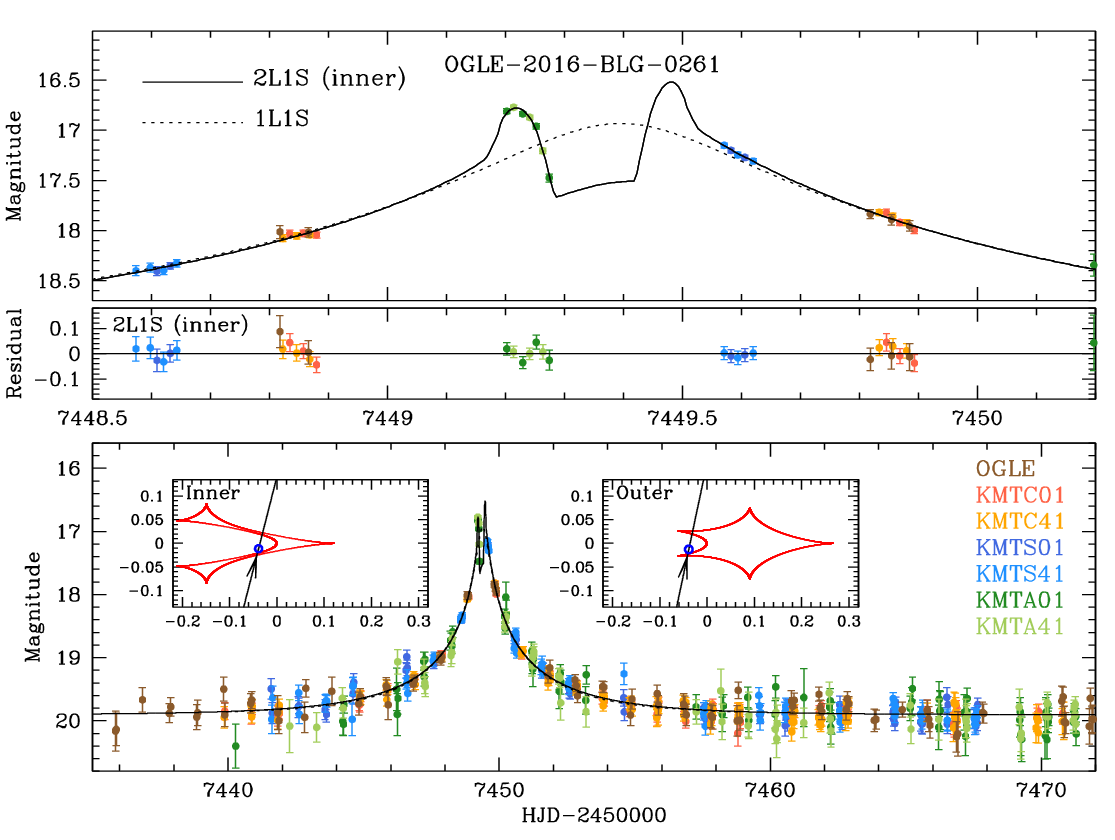}
\caption{
Light curve of the microlensing event OGLE-2016-BLG-0261. The bottom panel shows the full 
light curve, while the top panel presents a zoomed-in view of the peak region. The two insets 
in the bottom panel illustrate the lens-system configurations for the inner and outer 2L1S 
solutions.  The coordinate system is centered on the host star, and all lengths are scaled to 
the angular Einstein radius.  In each inset, the red cusp figure is the caustic and the arrowed 
line represents the source trajectory. The small empty blue circle at the tip of the arrow 
represents the source size scaled to Einstein radius. The solid and dotted curves overlaid on 
the data represent the model curves of the inner 2L1S and 1L1S models, respectively. The middle 
panel shows the residuals from the inner 2L1S model.
}
\label{fig:one}
\end{figure*}

To search for planetary solutions, we first perform a grid search over the binary lens parameters 
$(s, q, \alpha)$, which primarily determine the caustic structure, while optimizing the remaining 
parameters. This procedure enables the identification of local solutions arising from various 
types of degeneracy. Candidate solutions identified from the resulting $\chi^2$ surface are then 
refined by allowing all parameters to vary simultaneously.

In cases where the anomaly morphology exhibits a smooth, positive, short-term deviation, we 
additionally test single-lens binary-source (1L2S) models. A 1L2S model is described by the 
standard single-lens parameters of the primary source, together with additional parameters 
that characterize the secondary source, including the time of closest approach, $t_{0,2}$, 
impact parameter, $u_{0,2}$, normalized source radius, $\rho_2$ when finite-source effects 
are relevant, and the source flux ratio, $q_{\rm F}=F_2/F_1$, where $F_1$ and $F_2$ denote 
the fluxes of the primary and secondary source stars, respectively.  By contrast, the 2L1S 
model involves only a single source star with flux $F_s$. The baseline magnitude, $I_{\rm base}$, 
is the magnitude corresponding to the total baseline flux, $F_{\rm base}=F_s+F_b$, where $F_b$ 
denotes the blended flux.

Microlensing light-curve modeling often yields multiple degenerate solutions that provide similarly 
good fits to the data. Some of these degeneracies arise from intrinsic symmetries of the lensing 
geometry, such as the close--wide \citep{Griest1998, Dominik1999} and inner--outer \citep{Gaudi1997} 
degeneracies, while others can result from incomplete coverage of short-lived anomalous features. 
In addition, short-term positive anomalies may occasionally be mimicked by perturbations caused by 
a faint companion to the source rather than to the lens \citep{Gaudi1998, Gaudi2004}. In our analysis, 
we examine the nature of the degeneracies identified for each event. In the following subsections, 
we present the modeling results for the individual events and discuss the degeneracies encountered 
in each case.

\begin{deluxetable*}{l|ll|llll}
\tablewidth{0pt}
\tablecaption{Lensing parameters of OGLE-2016-BLG-0261 and KMT-2025-BLG-0026. \label{table:three}}
\tablehead{
\multicolumn{1}{c|}{Parameter}           &
\multicolumn{2}{c|}{OGLE-2016-BLG-0261}  &  
\multicolumn{3}{c}{KMT-2025-BLG-0026}  \\
\multicolumn{1}{c|}{ }                   &
\multicolumn{1}{c}{2L1S (inner)}        &
\multicolumn{1}{c|}{2L1S (outer)}        &            
\multicolumn{1}{c}{2L1S (close)}        &
\multicolumn{1}{c}{2L1S (wide)}         &
\multicolumn{1}{c}{1L2S } 
}
\startdata
 $\chi^2$                        & $5782.0              $  &  $5782.6              $    &  $1463.4             $  &  $1480.0             $   &  $1480.0             $    \\
 $t_0$ (HJD$^\prime$)            & $7449.4067 \pm 0.0044$  &  $7449.4032 \pm 0.0040$    &  $786.4955 \pm 0.0050$  &  $786.4945 \pm 0.0051$   &  $786.4900 \pm 0.0047$    \\
 $u_0$                           & $0.0345 \pm 0.0035   $  &  $0.0353 \pm 0.0035   $    &  $0.0928 \pm 0.0014  $  &  $0.0948 \pm 0.0016  $   &  $0.0936 \pm 0.0021  $    \\
 $\te$ (days)                    & $5.62 \pm 0.26       $  &  $5.55 \pm 0.26       $    &  $21.80 \pm 0.11     $  &  $21.57 \pm 0.11     $   &  $21.70 \pm 0.10     $    \\
 $s$                             & $0.928 \pm 0.016     $  &  $1.045 \pm 0.0145    $    &  $0.3976 \pm 0.0025  $  &  $2.5264 \pm 0.0166  $   &  \nodata                  \\
 $q$ ($10^{-3}$)                 & $2.76 \pm 0.66       $  &  $3.26 \pm 0.60       $    &  $4.62 \pm 0.90      $  &  $0.50 \pm 0.17      $   &  \nodata                  \\
 $\alpha$ (rad)                  & $1.820 \pm 0.068     $  &  $1.779 \pm 0.062     $    &  $3.2277 \pm 0.0118  $  &  $-0.02050 \pm 0.0044$   &  \nodata                  \\
 $\rho$ ($10^{-3}$)              & $7.47 \pm 0.63       $  &  $8.04 \pm 0.59       $    &  $34 \pm 13          $  &  $41 \pm 14          $   &  $< 70               $    \\
 $t_{0,2}$ ({\rm HJD}$^\prime$)  & \nodata                 &  \nodata                   &  \nodata                &  \nodata                 &  $740.653 \pm 0.327  $    \\
 $u_{0,2}$                       & \nodata                 &  \nodata                   &  \nodata                &  \nodata                 &  $0.036 \pm 0.019    $    \\
 $\rho_2$ ($10^{-3}$)            & \nodata                 &  \nodata                   &  \nodata                &  \nodata                 &  $< 2                $    \\
 $q_F$                           & \nodata                 &  \nodata                   &  \nodata                &  \nodata                 &  $0.0013 \pm 0.0003  $    \\
\enddata
\tablecomments{
 ${\rm HJD}^\prime\equiv {\rm HJD}-2450000$ for OGLE-2016-BLG-0261 and 
 ${\rm HJD}^\prime\equiv {\rm HJD}-2460000$ for KMT-2025-BLG-0026.
}
\end{deluxetable*}

\subsection{OGLE-2016-BLG-0261} \label{sec:three-one}

The lensing event OGLE-2016-BLG-0261 was observed by both OGLE and KMTNet surveys.  The source 
lies in the OGLE-IV bulge field BLG501, which is monitored at a cadence of $\sim 20$--$30$ min. 
Its position also falls in the overlapping region of the KMTNet prime fields BLG01 and BLG41, 
each observed with a cadence of 30 min, corresponding to an effective cadence of 15~min when 
combined. The source has a baseline magnitude of $I_{\rm base}=20.1$.  The extinction toward 
the field is $A_I = 2.66$, estimated using the Galactic bulge reddening maps of 
\citet{Gonzalez2012,Gonzalez2018}, which are based on VVV observations of red clump giants.  
The lensing magnification lasted for a relatively short duration, with an estimated event 
timescale of $\te \sim 5.6$~days.

Figure~\ref{fig:one} shows the lensing light curve constructed from the combined OGLE and KMTNet 
data. The event reached a relatively high magnification of $A_{\rm max}\sim 40$ at the peak at 
about ${\rm HJD}^\prime \equiv {\rm HJD} - 2450000 \sim 7449.5$. This region exhibits a deviation 
from the single-lens single-source (1L1S) model (dotted curve).  A possible anomaly, lasting 
approximately 2 hr, is present in the data processed by the automated photometry pipeline. It 
was subsequently confirmed through a reanalysis based on reprocessed data using the updated KMTNet 
photometry pipeline \citep{Yang2024}.  Deviations occurring near the peak of high-magnification 
events are usually associated with low-mass companions. This interpretation is further supported 
by the rapid variation in magnification, which is likely produced by a small caustic induced by 
such a companion.

\begin{figure*}[t]
\centering
\includegraphics[width=13.0cm]{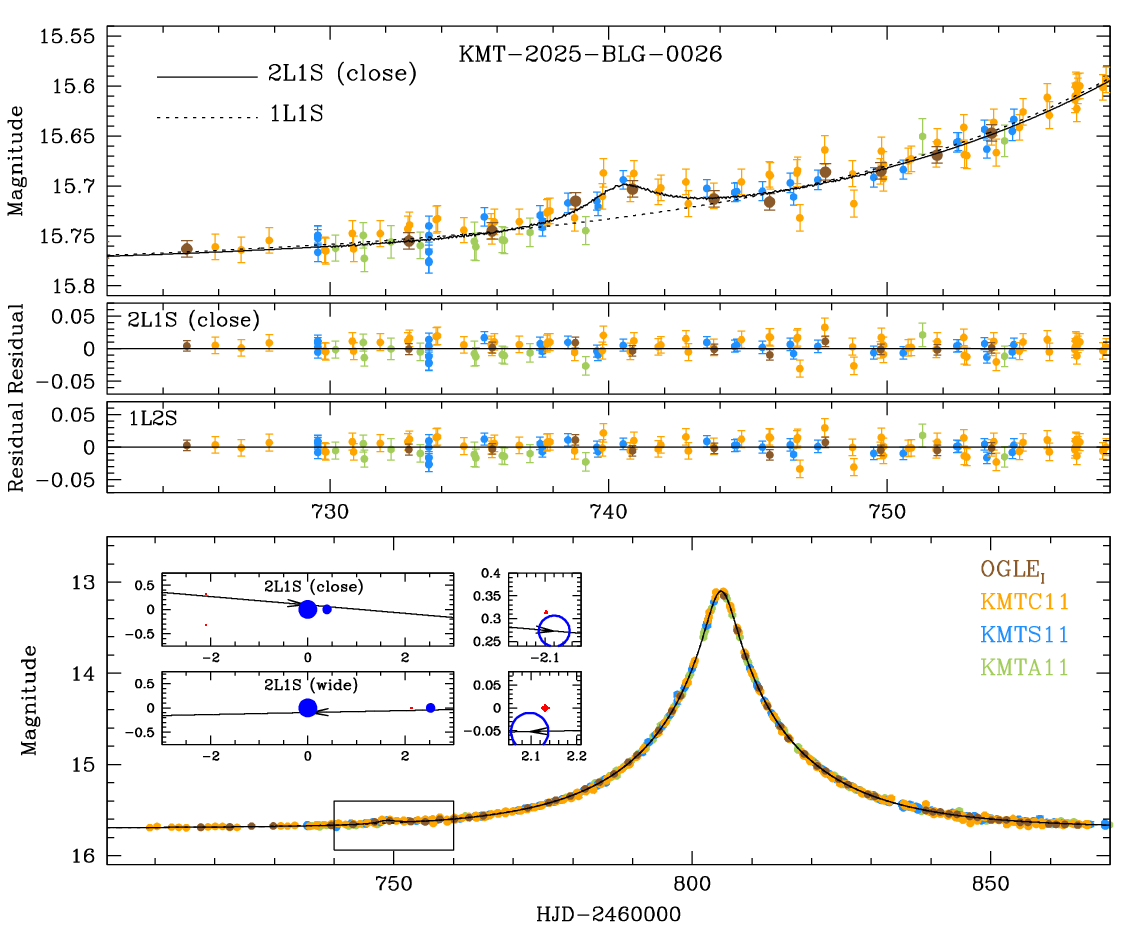}
\caption{
Lensing light curve of KMT-2025-BLG-0026.  The small box in the bottom panel marks the region 
of the anomaly. The two sets of insets in the bottom panel show the lens-system configurations 
of the close (upper inset) and wide (lower inset) solutions. For each solution, the left inset 
shows the source trajectory relative to the lens positions (marked by blue dots), and the right 
inset presents a zoomed-in view of the source near the peripheral caustic.
}
\label{fig:two}
\end{figure*}

Given the characteristics of the anomaly, we conduct a 2L1S modeling of the light curve, which
reveals a planetary solution that explains the observed perturbation. We identify two local
solutions: one with binary lens parameters $(s, q)\sim (0.93, 2.8\times 10^{-3})$ and the other
with $(s, q)\sim (1.05, 3.3\times 10^{-3})$. We designate these as the ``inner'' and ``outer''
solutions, respectively, for the reasons discussed below. For both solutions, the mass ratio 
between the binary lens component is of order $10^{-3}$, indicating the planetary origin of the 
anomaly.  The full set of lensing parameters for both solutions is listed in Table~\ref{table:three}. 
The two solutions are highly degenerate, with the inner solution favored by only a marginal difference 
of $\Delta\chi^2=0.6$. The model curve for the inner solution is shown in Figure~\ref{fig:one}. The 
normalized source radius is measured to be $\rho\sim (7\text{--}8)\times 10^{-3}$. 
Although the residuals exhibit a slight visual trend around ${\rm HJD}^\prime \sim 7448.83$, 
more complex models do not yield a statistically significant improvement to the fit.

The lens-system configurations corresponding to the inner and outer solutions are presented in 
the two insets of the bottom panel of Figure~\ref{fig:one}.  For both solutions, the lens companion 
induces a resonant caustic in which the central and peripheral caustics are merged.  In the inner 
solution, the source passes through the ``inner'' region between the central and peripheral caustics, 
whereas in the outer solution, the source traverses the ``outer'' region of the caustic.  Based on 
this relative geometry, we designate the solutions as ``inner'' and ``outer,'' respectively.  The 
anomaly arises as the source crosses two cusps located on the backside of the central caustic. The 
observed anomaly at approximately ${\rm HJD}^\prime \sim 7449.5$ corresponds to the first caustic 
crossing, while the second crossing is not covered by the available data.

The degeneracy between the two solutions is caused by the well-known inner--outer degeneracy.
\citet{Gaudi1997} first pointed out that this degeneracy arises between solutions with source
trajectories passing on the inner and outer sides of a peripheral caustic that is well separated
from the central caustic. Later, \citet{Yee2021} and \citet{Zhang2022} showed that a similar
degeneracy can also occur in planetary perturbations induced by both central and peripheral
caustics. \citet{Hwang2022} and \citet{Gould2022} derived analytic relations linking the
planetary separations of the inner ($s_{\rm in}$) and outer ($s_{\rm out}$) solutions:
\begin{equation}
\sqrt{s_{\rm in} \times s_{\rm out}} = s^\dagger;\qquad
s^\dagger =  {\sqrt{u_{\rm anom}^2 + 4} \pm u_{\rm anom} \over 2}.
\label{eq1}
\end{equation}
In this formulation, $u_{\rm anom} = (\tau_{\rm anom}^2 + u_0^2)^{1/2}$ is the lens--source 
separation at the time of the anomaly ($t_{\rm anom}$), where $\tau_{\rm anom} = (t_{\rm anom} 
- t_0)/t_{\rm E}$ is the corresponding normalized time offset.  The ``$+$'' sign applies to 
perturbations of the major image, which produce positive deviations from the underlying 1L1S 
light curve, whereas the ``$-$'' sign applies to perturbations of the minor image, which 
produce negative deviations.  For the pair of degenerate solutions identified in 
OGLE-2016-BLG-0261, the anomaly is a minor-image perturbation, and hence the ``$-$'' sign 
applies.  The resulting geometric mean of the planetary separations, $\sqrt{s_{\rm in}\times 
s_{\rm out}}\sim0.98$, agrees well with the predicted value of $s^\dagger$.

\subsection{KMT-2025-BLG-0026} \label{sec:three-two}

The microlensing event KMT-2025-BLG-0026 was observed by the KMTNet and OGLE surveys. The source 
lies in subfields of both surveys that are monitored at relatively low cadence. It has a baseline 
magnitude of $I_{\rm base} = 15.76$, suggesting a bulge giant if most of the baseline flux originates 
from the lensed star. The extinction toward the field is $A_I = 1.81$.

Figure~\ref{fig:two} shows the lensing light curve of the event. At first glance, the light 
curve appears to be that of a typical 1L1S event.  However, a detailed inspection reveals a 
weak anomaly (of $\sim 0.04$ mag) around ${\rm HJD}^\prime \equiv {\rm HJD}-2460000 \sim 741$. 
We first verified the genuineness of the signal by re-reducing the data and then by checking the 
consistency between the OGLE and KMTNet data sets. From these tests, we confirm that the signal 
is real. The zoomed-in view of the region of the anomaly is shown in the top panel.

\begin{figure*}[t]
\centering
\includegraphics[width=13.0cm]{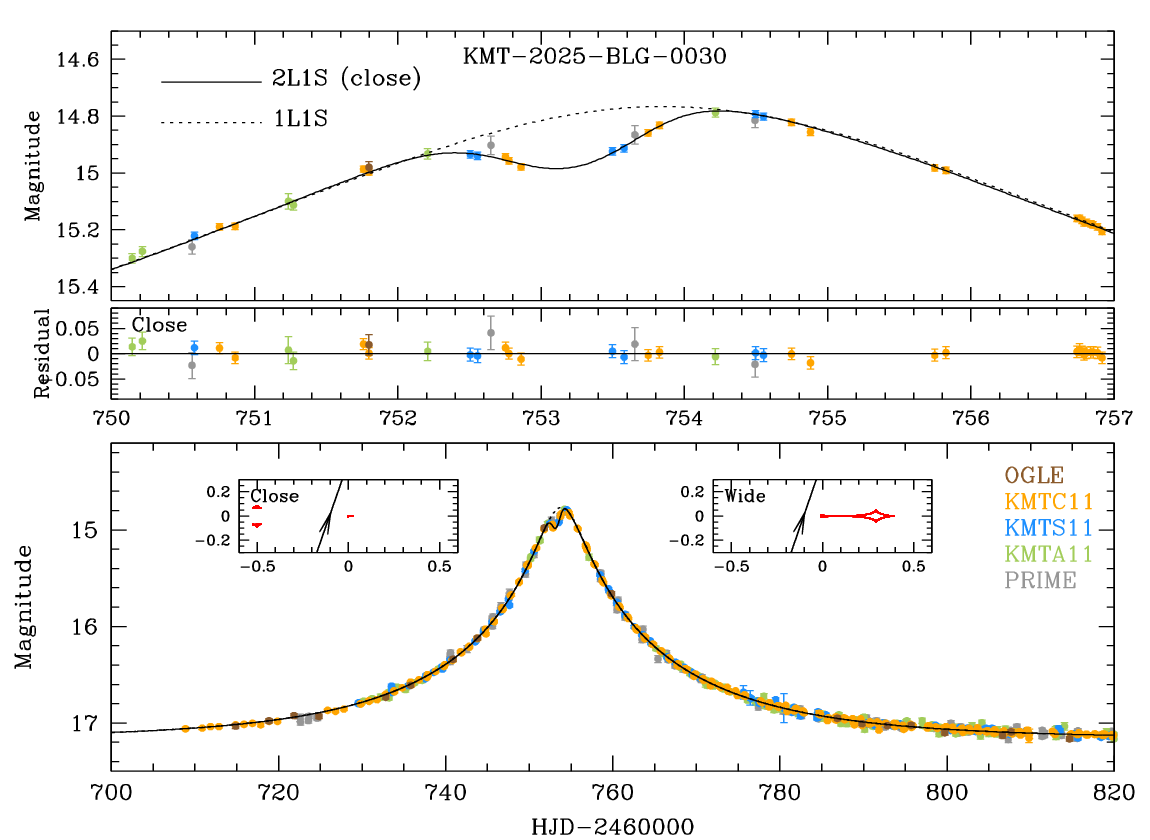}
\caption{
Light curve of lensing event KMT-2025-BLG-0030.
The notation and layout are the same as those in Fig.~\ref{fig:one}.
}
\label{fig:three}
\end{figure*}

We initially modeled the anomaly using a 2L1S configuration. From this, we identified two local
solutions: one in the close-binary regime ($s<1$) and the other in the wide-binary regime ($s>1$).
The close solution has binary parameters $(s, q)\sim (0.40, 4.6\times 10^{-3})$, while the wide
solution has $(s, q)\sim (2.53, 0.5\times 10^{-3})$. The low mass ratios indicate that the anomaly
is produced by a planetary companion. Notably, while the planet separations of the two solutions
approximately follow the $s_{\rm c}\sim 1/s_{\rm w}$ relation, the mass ratio of the close solution 
is roughly an order of magnitude higher than that of the wide solution. The event timescale is 
estimated to be $\te\sim 22$~days. The complete lensing parameters of both solutions are presented 
in Table~\ref{table:three} together with the $\chi^2$ values of the fits. We find that the close 
solution is favored over the wide solution by $\Delta\chi^2=16.6$. In Figure~\ref{fig:two}, we 
present the model curve and residuals of the close solution.

We additionally tested a 1L2S interpretation of the anomaly. The corresponding lensing
parameters are presented in Table~\ref{table:three}. In this model, the anomaly is attributed 
to a very faint companion source with a flux ratio of $q_F \sim 0.001$ relative to the primary 
and leading it by a time offset of $\Delta t \sim 46$ days.  Given that the primary source is 
a giant star (as discussed in Sect.~\ref{sec:four}), the secondary source would be a K dwarf 
with a similar color. Its radius is therefore expected to be smaller than that of the primary 
by a factor of $\sim \sqrt{q_F}\simeq 30$, and we accordingly impose the constraint $\rho_2 < 
0.002$ in the modeling.  The residuals of this fit are shown in Figure~\ref{fig:two}. 
Although the 1L2S model provides a fit comparable to that of the wide 2L1S solution, it is 
disfavored with respect to the close 2L1S solution by $\Delta\chi^2=16.6$.

\begin{deluxetable*}{lllllll}
\tablewidth{0pt}
\tablecaption{Lensing parameters KMT-2025-BLG-0030 and KMT-2025-BLG-2272.  \label{table:four}}
\tablehead{
\multicolumn{1}{c}{Parameter}           &
\multicolumn{2}{c}{KMT-2025-BLG-0030}   &
\multicolumn{2}{c}{KMT-2025-BLG-2272}   \\
\multicolumn{1}{c}{ }                   &
\multicolumn{1}{c}{Close}               &
\multicolumn{1}{c}{Wide }               &
\multicolumn{1}{c}{Inner}               &
\multicolumn{1}{c}{Outer}       
}
\startdata
 $\chi^2$                     &  $1378.6             $ &  $1381.2             $   &  $8287.1          $   &  $8296.5           $ \\
 $t_0$ ({\rm HJD}$^\prime$)   &  $753.8831 \pm 0.0090$ &  $753.8758 \pm 0.0097$   &  $920.980 \pm 0.26$   &  $921.162 \pm 0.268$ \\
 $u_0$                        &  $0.1006 \pm 0.0016  $ &  $0.1002 \pm 0.0016  $   &  $0.493 \pm 0.019 $   &  $0.469 \pm 0.019  $ \\
 $\te$ (days)                 &  $28.62 \pm 0.27     $ &  $28.68 \pm 0.28     $   &  $46.73 \pm 1.44  $   &  $49.45 \pm 1.72   $ \\
 $s$                          &  $0.779 \pm 0.028    $ &  $1.157 \pm 0.040    $   &  $1.345 \pm 0.009 $   &  $1.219 \pm 0.015  $ \\
 $q$ ($10^{-3}$)              &  $1.84 \pm 0.26      $ &  $1.84 \pm 0.26      $   &  $3.99 \pm 0.37   $   &  $3.32 \pm 0.29    $ \\
 $\alpha$ (rad)               &  $1.7993 \pm 0.0065  $ &  $1.7969 \pm 0.0063  $   &  $4.954 \pm 0.011 $   &  $4.959 \pm 0.011  $ \\
 $\rho$ ($10^{-3}$)           &  $ < 30              $ &  $< 30               $   &  $1.832 \pm 0.084 $   &  $1.547 \pm 0.074  $ \\
\enddata
\tablecomments{
 ${\rm HJD}^\prime\equiv {\rm HJD}-2460000$.
}
\end{deluxetable*}

The two sets of insets in the bottom panel of Figure~\ref{fig:two} illustrate the lens-system
configurations for the close and wide 2L1S solutions. In the close solution, the planet induces 
two peripheral caustics located on opposite sides of the planet, whereas the wide solution features 
a single peripheral caustic on the same side as the planet. In both cases, the anomaly is produced 
by the source passing over a peripheral caustic. Because the source size is substantially larger 
than the caustic, the anomaly is strongly affected by finite-source effects. In this regime, the 
duration of the anomaly is determined by the source size, $\Delta t \sim 2\rho \te \sim 2$~days, 
which is consistent with the observations. Furthermore, the amplitude of the anomaly is governed 
by the ratio of the caustic size to the normalized source radius \citep{Bennett1996}.  Since the 
size of the peripheral caustic scales as $\sqrt{q}$ and is substantially larger in the wide solution 
than in the close solution \citep{Han2006}, the close solution requires a higher mass ratio to 
reproduce an anomaly with an amplitude comparable to that of the wide solution.  This explains 
why the estimated mass ratio of the close solution is greater than that of the wide solution.

Although we designate the solutions as ``close'' and ``wide,'' the similarity between their 
model light curves does not correspond to the degeneracy identified by \citet{Griest1998}. 
In the classical close--wide degeneracy, both solutions involve perturbations of the same 
image; that is, either both correspond to major-image perturbations or both to minor-image 
perturbations. In contrast, for KMT-2025-BLG-0026, the ``close'' solution arises from a 
minor-image perturbation, whereas the ``wide'' solution arises from a major-image perturbation. 
The degeneracy is therefore accidental, in the sense that it does not originate from an 
underlying symmetry of the lens equation.

\subsection{KMT-2025-BLG-0030} \label{sec:three-three}

The lensing event KMT-2025-BLG-0030 was observed by the three surveys KMTNet, OGLE, and PRIME.
The source is located in the KMTNet subfield BLG11, which was monitored with a cadence of 2.5~hours.
The extinction toward the field is relatively high, with $A_I = 2.94$. The baseline magnitude of 
the source is $I_{\rm base} = 17.81$, suggesting that the source is likely a giant star.

The lensing light curve of the event is presented in Figure~\ref{fig:three}. The event reached 
its peak at ${\rm HJD}^\prime \sim 753.9$ with a moderately high magnification of $A_{\rm max} 
\sim 10$. Shortly before the peak, the light curve exhibits an anomaly characterized by a dip 
feature and a deviation of approximately 0.2 mag from the underlying 1L1S model. This anomaly 
persists for about 1.5 days. The Einstein timescale of the event is estimated to be $\te \sim 29$ 
days, based on a 1L1S model that excludes the data points surrounding the anomaly.

\begin{figure*}[t]
\centering
\includegraphics[width=13.0cm]{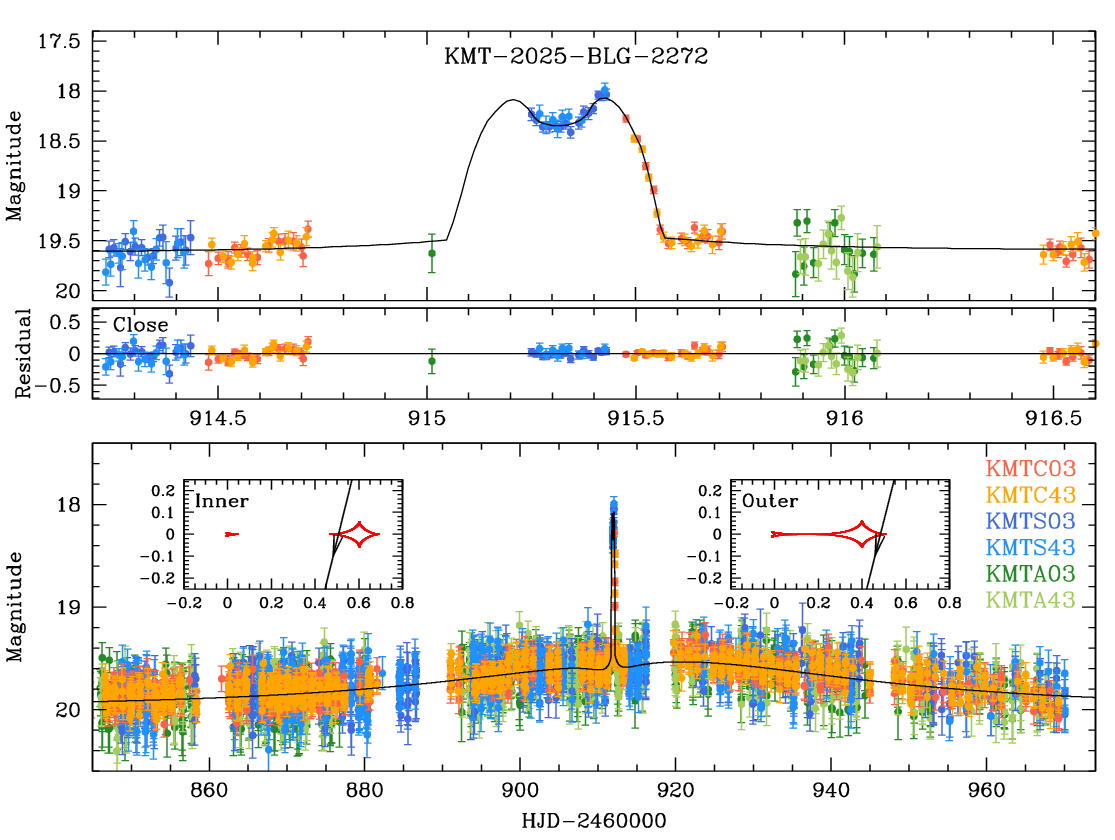}
\caption{
Light curve of KMT-2025-BLG-2272. 
The figure follows the same format as Fig.~\ref{fig:one}.
}
\label{fig:four}
\end{figure*}

Because a low-mass companion to the lens is a well-known origin for a short-term negative deviation, 
we modeled the light curve using a 2L1S configuration. From this analysis, we identified a pair 
of planetary solutions in the close and wide regimes with parameters $(s, q) \sim (0.78, 1.8
\times 10^{-3})$ for the close solution and $(s, q) \sim (1.16, 1.8 \times 10^{-3})$ for the wide
solution. We find that the close solution is favored by $\Delta\chi^2 = 2.6$. The complete sets 
of lensing parameters for both solutions are presented in Table~\ref{table:four}. The model curve 
representing the close solution is overlaid on the data points in Figure~\ref{fig:three}.

The configurations of the lens system for both the close and wide solutions are illustrated in the
two insets within the bottom panel of Figure~\ref{fig:three}. For the close solution, the planetary
companion induces a central caustic along with a pair of peripheral caustics on the opposite side,
whereas for the wide solution, it induces a central caustic and a single peripheral caustic on the
planet side. The anomaly observed near the peak magnification was produced as the source passed
the backside of the central caustic. Because the source-caustic separation during the anomaly was
considerable, the normalized source radius could not be measured and only its upper limit of
$\rho_{\rm max} \sim 0.03$ is constrained.

\subsection{KMT-2025-BLG-2272} \label{sec:three-four}

The lensing magnification of KMT-2025-BLG-2272 was detected and observed exclusively by the
KMTNet survey. The source star is located in the overlapping region between the KMTNet prime
fields BLG03 and BLG43, toward which observations were conducted with a combined cadence of
15 minutes. The extinction toward the field is $A_I = 2.17$. Because the source is very faint and
the lensing magnification at the peak was low, the existence of an anomaly was difficult to notice
from the online data and could only be securely revealed through a rigorous re-reduction of the
data.

Figure~\ref{fig:four} shows the lensing light curve of the event.  It exhibits a brief anomaly 
around ${\rm HJD}^\prime \sim 915.3$, when the underlying 1L1S magnification is $A \sim 2$.  
The anomaly lasts for approximately half a day.  Its profile, characterized by a U-shaped feature 
followed by a sharp drop, is indicative of a caustic-crossing event.  The event timescale, 
estimated from a 1L1S fit to the light curve excluding the anomalous data, is $\te \sim 45$~days.

\begin{figure*}[t]
\centering
\includegraphics[width=15.0cm]{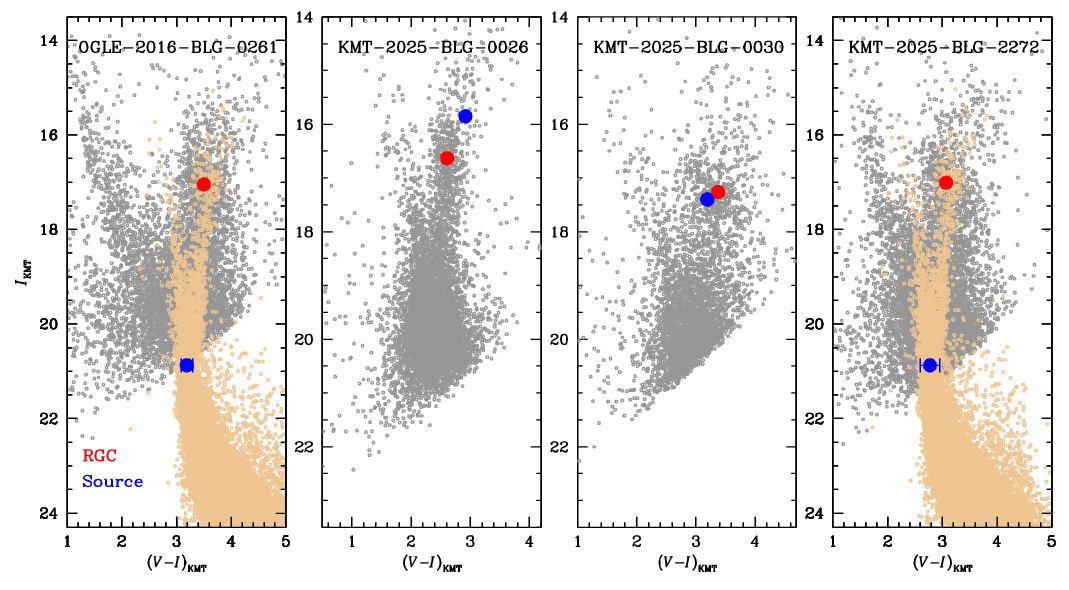}
\caption{
Locations of the source stars relative to the centroids of the red giant clump (RGC) in the 
color--magnitude diagram (CMD) constructed from KMTC observations (gray dots).  For events 
whose source colors are determined using the CMD from HST observations, the HST CMD (brown 
dots) is also shown.
}
\label{fig:five}
\end{figure*}

From a 2L1S modeling conducted with the consideration of the observed caustic features, we 
identify a pair of solutions resulting from the inner--outer degeneracy. The binary parameters 
are $(s, q) \sim (1.35, 4.0 \times 10^{-3})$ for the inner solution and $(s, q) \sim (1.22, 3.3 
\times 10^{-3})$ for the  outer solution. The low mass ratio indicates that the companion to the 
lens is a planetary-mass object. The normalized source radius was securely measured from the 
resolved caustic features during the anomaly. The full lensing parameters for both solutions 
are listed in Table~\ref{table:four}. Among the two models, the inner solution is preferred 
over the outer solution by $\Delta\chi^2 = 9.4$. The model curve of the inner solution and its 
residuals are presented in Figure~\ref{fig:four}.  As expected from the low peak magnification 
of the underlying 1L1S event, the lens--source impact parameter is relatively large, with 
$u_0 \sim 0.47$--0.49.

The two insets in the bottom panel of Figure~\ref{fig:four} show the configurations of the inner 
and outer solutions. They indicate that the anomaly is produced as the source passes near a peripheral 
caustic. In the inner solution, the source traverses the inner region of the caustic, whereas in the 
outer solution, it passes through the outer region. This demonstrates that the degeneracy between 
the two solutions is a typical example of the inner--outer degeneracy discussed by \citet{Gaudi1997}. 
The anomaly occurs near the peak of the underlying 1L1S light curve because the source crosses the 
caustic at an incidence angle ($\sim 76^\circ$) close to a right angle.

\begin{deluxetable*}{llllllllll}
\tablewidth{0pt}
\tablecaption{Source parameters, angular Einstein radius, and relative lens-source proper motion.\label{table:five}}
\tablehead{
\multicolumn{1}{c}{Event}                    &
\multicolumn{1}{c}{$(V-I, I)_0$}             &
\multicolumn{1}{c}{Type}                     &
\multicolumn{1}{c}{$\theta_*$ ($\mu$as)}     &
\multicolumn{1}{c}{$\thetae$ (mas)}          & 
\multicolumn{1}{c}{$\mu$ (mas/yr)}                
}
\startdata
OGLE-2016-BLG-0261 &  $(0.749 \pm 0.105, 18.307 \pm 0.089)$ &   G5V    &   $0.718 \pm 0.091$   &   $0.089 \pm 0.015$   &   $6.18 \pm 1.14$   \\
KMT-2025-BLG-0026  &  $(1.369 \pm 0.040, 13.789 \pm 0.020)$ &   K3III  &   $10.21 \pm 0.83 $   &   $0.302 \pm 0.118$   &   $5.06 \pm 1.99$   \\
KMT-2025-BLG-0030  &  $(0.877 \pm 0.042, 14.563 \pm 0.020)$ &   G6III  &   $4.634 \pm 0.379$   &   $> 0.15         $   &   $> 1.9        $   \\
KMT-2025-BLG-2272  &  $(0.764 \pm 0.181, 18.230 \pm 0.018)$ &   G7V    &   $0.756 \pm 0.147$   &   $0.413 \pm 0.082$   &   $3.23 \pm 0.65$   \\
\enddata
\tablecomments{
$(V-I, I)_0$ denote the de-reddened color and magnitude of the source star, ``Type'' denotes 
the spectral type of the source, $\theta_*$ is the angular source radius, $\thetae$ is the 
angular Einstein radius, and $\mu$ is the relative lens--source proper motion.
 }
\end{deluxetable*}

\section{Angular Einstein radii} \label{sec:four}

The physical parameters of the lens mass and distance can be constrained by lensing observables,
including the event timescale, angular Einstein radius, and microlens parallax ($\pie$). The 
event timescale is measured from light-curve modeling.  In principle, the microlens parallax can 
be determined from subtle deviations in the lensing light curve from a static model caused by the 
Earth's orbital motion around the Sun \citep{Gould1992}.  However, for none of the analyzed events 
can $\pie$ be measured, either because the photometric precision is insufficient to detect such 
deviations or because the event durations are not sufficiently long.

The other lensing observable of the angular Einstein radius can be measured for events with a 
well-determined normalized source radius via the relation
\begin{equation}
\thetae = \frac{\theta_*}{\rho},
\label{eq2}
\end{equation}
\hskip-4pt
where $\theta_*$ denotes the angular radius of the source star. For  three out of the four events 
analyzed in this work, the normalized source radius is securely measured. In this section, we 
estimate the angular Einstein radii for these events.

The angular source radius is inferred from the reddening-corrected color and magnitude of the
source. We derive the de-reddened source color and magnitude, $(V - I, I)_0$, by applying a
color--magnitude diagram (CMD) calibration relative to the centroid of the red giant clump 
(RGC) in the same field, following the method of \citet{Yoo2004}. The RGC centroid provides 
an excellent reference because it lies behind nearly the same column of interstellar dust as 
the source, and its intrinsic color and magnitude are well established for the Galactic bulge 
field \citep{Bensby2013, Nataf2013}.

Figure~\ref{fig:five} shows the positions of the source stars relative to the RGC centroids for 
the individual events.  For the two events OGLE-2016-BLG-0261 and KMT-2025-BLG-2272, the source 
stars are too faint to securely measure their instrumental $V$-band magnitudes.  In these cases, 
we align the CMD constructed from KMTC observations with that derived from Hubble Space Telescope 
(HST) observations \citep{Holtzman1998}, and then infer the source color as the mean color of HST 
stars that lie within the measured range of $I$-band magnitude offsets from the RGC centroid.

Table~\ref{table:five} lists the obtained de-reddened color and magnitude of the source stars 
together with their spectral types. We find that the source stars of KMT-2025-BLG-0026 and 
KMT-2025-BLG-0030 are giant stars with spectral types K3III and G6III, respectively. The source 
stars of the other events are main-sequence stars with spectral types of G5V for OGLE-2016-BLG-0261
and G7V for KMT-2025-BLG-2272.

Using the de-reddened color and magnitude, we derive the angular source radius by adopting 
the surface-brightness relations of \citet{Kervella2004}. Because this relation provides the 
correspondence between $(V-K, V)$ and $\theta_*$, we initially transform $V-I$ to $V-K$ 
using the color--color relations of \citet{Bessell1988}.  With the estimated angular source 
radius, the angular Einstein radius is derived using Equation~(\ref{eq2}), and the relative 
lens--source proper motion is then calculated as $\mu = \theta_{\rm E}/t_{\rm E}$.  
Table~\ref{table:five} lists the estimated values of $\theta_*$, $\thetae$, and $\mu$ for the 
events. In the case of KMT-2025-BLG-0030, for which only the upper limit of $\rho$ is constrained, 
we provide the corresponding lower limits for $\thetae$ and $\mu$.

The values of $\theta_{\rm E}$ are derived using the $\rho$ values of the adopted solutions. For 
events affected by intrinsic degeneracies (e.g., the inner--outer or close--wide degeneracies), 
the degenerate solutions yield similar values of $\rho$, and therefore the derived values of 
$\thetae$ are similar regardless of which solution is adopted. KMT-2025-BLG-0026 is an exception 
because the close and wide solutions arise from an accidental degeneracy and have significantly 
different values of $\rho$. For this event, we adopt the close solution, which is favored over 
the wide solution by $\Delta\chi^2 = 16.6$, to derive $\thetae$.

\begin{deluxetable*}{llccccccc}
\tablewidth{0pt}
\tablecaption{Physical lens parameters.\label{table:six}}
\tablehead{
\multicolumn{1}{c}{Event}                                 &
\multicolumn{1}{c}{Solution}                              &
\multicolumn{1}{c}{$M_{\rm h}$ ($M_\odot$)}               &
\multicolumn{1}{c}{$M_{\rm p}$ ($M_{\rm J}$)}             &
\multicolumn{1}{c}{$\dl$ (kpc)}                           &
\multicolumn{1}{c}{$a_\perp$ (au)}                        & 
\multicolumn{1}{c}{$p_{\rm disk}$ }                       &
\multicolumn{1}{c}{$p_{\rm bulge}$}  
}
\startdata
OGLE-2016-BLG-0261 & Inner   &  $0.072^{+0.117}_{-0.040}$  &  $0.21^{+0.34}_{-0.12}$   &   $7.85^{+0.97}_{-1.07}$  &  $0.72^{+0.09}_{-0.10}$   &   21\%   &    79\%  \\  [0.8ex]
                   & Outer   &                             &  $0.25^{+0.40}_{-0.14}$   &                           &  $0.81^{+0.10}_{-0.11}$   &          &          \\  [0.8ex]
\hline                                                                                                                                                                 
KMT-2025-BLG-0026  & Close   &  $0.44^{+1.08}_{-0.26}   $  &  $2.13^{+5.23}_{-1.26}$   &   $7.28^{+1.10}_{-1.53}$  &  $0.96^{+0.15}_{-0.20}$   &   30\%   &    70\%  \\  [0.8ex]
                   & Wide    &  $0.23^{+0.57}_{-0.14}   $  &                           &                           &  $6.06^{+0.92}_{-1.27}$   &          &          \\  [0.8ex]
\hline                                                                                                                                                                 
KMT-2025-BLG-0030  & Close   &  $0.60^{+0.36}_{-0.34}   $  &  $1.15^{+0.69}_{-0.66}$   &   $6.72^{+1.26}_{-2.16}$  &  $2.35^{+0.44}_{-0.76}$   &   45\%   &    55\%  \\  [0.8ex]
                   & Wide    &                             &                           &                           &  $3.49^{+0.66}_{-1.12}$   &          &          \\  [0.8ex]
\hline                                                                                                                                                               
KMT-2025-BLG-2272  & Inner   &  $0.63^{+0.31}_{-0.32}   $  &  $2.47^{+1.23}_{-1.25}$   &   $6.62^{+0.90}_{-1.32}$  &  $3.89^{+0.53}_{-0.78}$   &   37\%   &    63\%  \\  [0.8ex]
                   & Outer   &                             &  $2.22^{+1.11}_{-1.12}$   &                           &  $3.51^{+0.48}_{-0.70}$   &          &          \\  [0.8ex]         
\enddata
\end{deluxetable*}

\begin{figure}[t]
\includegraphics[width=\columnwidth]{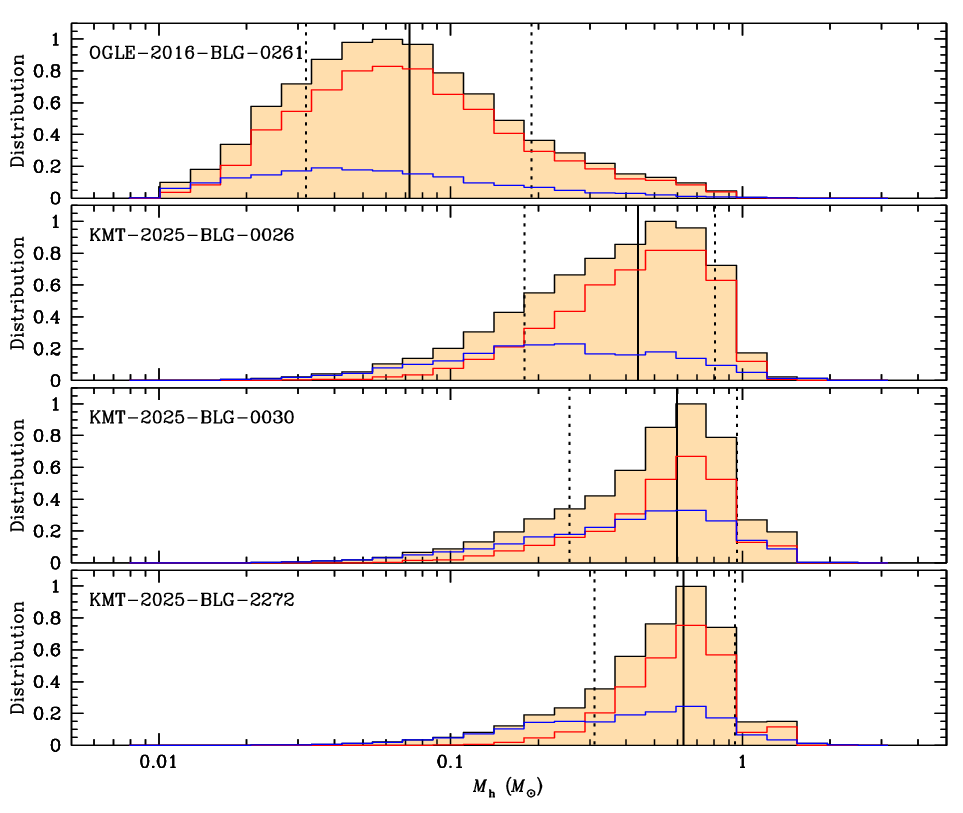}
\centering
\caption{
Bayesian posterior distributions for the host masses of the planetary systems. In each panel, the
solid vertical line represents the median value, and the two dotted vertical lines indicate the $1\sigma$
uncertainty range. The blue and red curves show the contributions from the disk and bulge lens
populations, respectively, while the black curve represents the sum of the two populations.
}
\label{fig:six}
\end{figure}

\section{Physical lens parameters} \label{sec:five}

The lens mass ($M$) and distance ($\dl$) are estimated through a Bayesian analysis that utilizes
the constraints provided by the measured observables of $\te$ and $\thetae$:
\begin{equation}
\te = \frac{\thetae}{\mu};\qquad
\thetae = \sqrt{\kappa M \pi_{\rm rel}}.
\label{eq3}
\end{equation}
\hskip-3pt
Here $\kappa = 4G/(c^2{\rm au}) \simeq 8.144~{\rm mas}~M_\odot^{-1}$, 
$\pi_{\rm rel} = \pi_{\rm L} - \pi_{\rm S} = {\rm au}(1/\dl – 1/\ds)$ is the relative lens-source 
parallax, and $\ds$ denotes the distance to the source.

The Bayesian analysis begins by generating a large ensemble of artificial microlensing events using 
a Monte Carlo simulation. In this process, the physical properties of lenses and sources are drawn 
from a Galactic model that specifies the spatial density distributions of the disk and bulge populations, 
their kinematics, and an underlying lens mass function. For each simulated event, the lens mass $M$, 
lens distance $\dl$, and source distance $\ds$ are assigned, and the relative transverse velocity is 
used to compute the relative proper motion $\mu$. From these quantities, the corresponding lensing 
observables, including $\te$ and $\thetae$, are derived using the relations in Equation~(\ref{eq3}). 
In our analysis, we adopt the Galactic model presented by \citet{Jung2021} and the lens mass-function 
prescription described by \citet{Jung2022}.

\begin{figure}[t]
\includegraphics[width=\columnwidth]{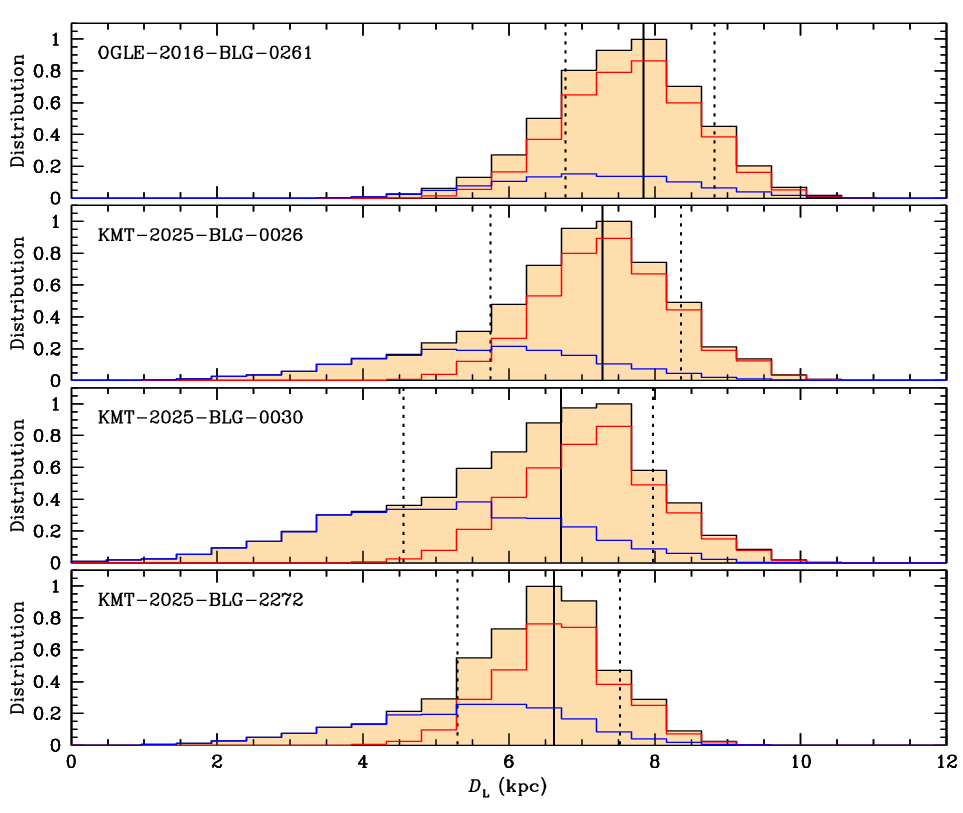}
\centering
\caption{
Bayesian posterior distributions for the
distances to the planetary systems. The notation is the same as in Fig.~\ref{fig:six}.
}
\label{fig:seven}
\end{figure}

Each simulated event is assigned a likelihood weight, $w_i$, according to the degree of agreement 
between its predicted observables and the measured values. This is implemented by evaluating a 
Gaussian likelihood function based on the differences between the simulated and observed $\te$ 
and $\thetae$, while accounting for their respective uncertainties:
\begin{equation}
w_i \propto \exp\left(-\frac{\chi^2}{2}\right),
\label{eq4}
\end{equation}
with
\begin{equation}
\chi^2 =
\left[ \frac{t_{{\rm E},i} - t_{{\rm E,obs}}}{\sigma(\te)} \right]^2 +
\left[ \frac{\theta_{{\rm E},i} - \theta_{{\rm E,obs}}}{\sigma(\thetae)} \right]^2.
\label{eq5}
\end{equation}
\hskip-4pt
Here, $(t_{\rm E,obs}, \theta_{\rm E,obs})$ denote the measured values of the observables, and
$[\sigma(\te), \sigma(\thetae)]$ represent their associated uncertainties.

The weighted ensemble of simulated events is used to construct posterior probability distributions
distributions of the host mass and the distance to the lens system. The estimated host mass 
($M_{\rm h}$), planet mass ($M_{\rm p}$), lens distance ($\dl$), and projected physical separation 
($a_\perp = s\dl\thetae$) between the planet and host are summarized in Table~\ref{table:six}. For 
each parameter, the median of the posterior distribution is adopted as the representative value, 
while the 16th and 84th percentiles are taken as the $1\sigma$ confidence intervals. The table also 
lists the relative contributions of disk ($p_{\rm disk}$) and bulge ($p_{\rm bulge}$) lenses, 
determined from the fraction of weighted events associated with each population.

The inferred companion masses lie predominantly in the giant-planet regime, typically in the range 
$\sim 0.2$--$2.5~M_{\rm J}$. The host masses span low- to intermediate-mass stars, with the exception 
of OGLE-2016-BLG-0261L, whose host lies near the boundary between a brown dwarf and a star. The lens 
distances cover a range from $\sim 6.6$ to $\sim 7.9$~kpc, with all systems more likely located in 
the Galactic bulge.  The projected separations are typically $\sim 1$--$6$ au, indicating that the 
planets lie at or beyond the snow line of their host stars. Overall, these results suggest that the 
detected companions are predominantly giant planets orbiting low-mass stars, consistent with the 
cold planet population probed by microlensing.

\section{Summary and conclusion} \label{sec:six}

We presented the discovery and analysis of four cold giant planets detected in the microlensing
events OGLE-2016-BLG-0261, KMT-2025-BLG-0026, KMT-2025-BLG-0030, and KMT-2025-BLG-2272. The 
planetary signals were identified from high-cadence survey data obtained by KMTNet, OGLE, and 
PRIME. In all cases, the observed anomalies were successfully explained by 2L1S planetary models, 
and the inferred mass ratios, typically of order $q\sim 10^{-3}$, indicate that the companions 
are giant planets.

Among the four events, two (OGLE-2016-BLG-0261 and KMT-2025-BLG-0030) exhibit anomalies associated 
with central caustics, while two events (KMT-2025-BLG-0026 and KMT-2025-BLG-2272) are explained by 
source passages over or near peripheral caustics.  Two events (OGLE-2016-BLG-0261 and KMT-2025-BLG-2272) 
show inner--outer degeneracies, whereas KMT-2025-BLG-0030 exhibits the classical close--wide degeneracy, 
and KMT-2025-BLG-0026 exhibits an accidental degeneracy.  For KMT-2025-BLG-0026, we also tested 1L2S 
interpretations, but these are disfavored relative to the planetary models. These detections therefore 
add four robust planetary systems to the sample of cold planets discovered by microlensing surveys.

Finite-source effects are securely measured for three events, excluding KMT-2025-BLG-0030, enabling 
determinations of the angular Einstein radius.  The measured values span a broad range, from 
$\theta_{\rm E} \sim 0.09$ mas for OGLE-2016-BLG-0261 to $\theta_{\rm E} \sim 0.41$ mas for 
KMT-2025-BLG-2272.  The corresponding relative lens--source proper motions are typically 
a few mas~yr$^{-1}$.

A Bayesian analysis based on the measured $\te$ and $\thetae$ indicates that the companions lie
predominantly in the giant-planet regime, with typical masses of $\sim 0.2$--$2.5~M_{\rm J}$.
The host masses are inferred to range from very low-mass to intermediate-mass stars, $\sim
0.07$--$0.63~M_\odot$, although the host of OGLE-2016-BLG-0261 is located near the stellar/substellar 
boundary. The projected separations are typically $\sim 0.7$--$6$ au, placing all of the planets 
beyond, or at least comparable to, the snow line of their host systems.  The inferred lens distances 
cover a range from (6.6--7.9) kpc, with most systems consistent with bulge lenses.

These results expand the sample of cold giant planets detected from homogeneous high-cadence
survey data and illustrate the continuing power of microlensing to probe planetary systems
inaccessible to the radial-velocity and transit methods. The four planets presented here 
reinforce the view that giant planets are present around low-mass hosts at separations of a 
few au and provide additional empirical constraints on the demographics of planets beyond the 
snow line. As the sample of uniformly analyzed survey planets continues to grow, such events 
will play an increasingly important role in clarifying the dependence of giant-planet occurrence 
on host mass and Galactic environment.

Several of the analyzed events exhibit close--wide or inner--outer degeneracies, although 
these generally have little effect on the inferred physical properties of the planetary 
systems. Future high-angular-resolution imaging, obtained with facilities such as the 
{\it Hubble Space Telescope}, the {\it James Webb Space Telescope}, extremely large 
telescopes, or space missions including {\it Roman} and {\it Euclid}, may help resolve 
these degeneracies. By measuring the lens flux and the lens--source relative proper motion 
after the source and lens have separated sufficiently, it may be possible to distinguish 
between the competing solutions and directly constrain the host masses and distances.

\begin{acknowledgements}
C.H. was supported by the Chungbuk National University 2025 NUDP program and the National 
Research Foundation of Korea (RS-2025-21073000).
This research has made use of the KMTNet system operated by the Korea Astronomy and Space 
Science Institute (KASI) at three host sites of CTIO in Chile, SAAO in South Africa, and 
SSO in Australia. Data transfer from the host site to KASI was supported by the Korea 
Research Environment Open NETwork (KREONET). This research was supported by KASI under 
the R\&D program (project No. 2026-1-904-01) supervised by the Ministry of Science and ICT.
W.Z. and H.Y. acknowledge support by the National Natural Science Foundation of China 
  (Grant No. 12133005). 
H.Y. acknowledges support by the China Postdoctoral Science Foundation (No. 2024M762938). 
W.Zang acknowledges the support from the Harvard-Smithsonian Center for Astrophysics 
  through the CfA Fellowship. 
J.C.Y. acknowledges support from U.S. NASA Grant No. 80NSSC25K7146. 
J.C.Y. acknowledges support from a Scholarly Studies grant from the Smithsonian Institution.
The OGLE project has received funding from the Polish National Science
Centre grant OPUS-28 2024/55/B/ST9/00447 awarded to AU.
The MOA/PRIME project is supported by JSPS KAKENHI Grant Number JP16H06287, JP22H00153 and 23KK0060.
C.R. was supported by the Research fellowship of the Alexander von Humboldt Foundation.
The PRIME project is supported by JSPS KAKENHI Grant Number JP16H06287, JP22H00153, JP25H00668,
JP19KK0082, JP20H04754, JP24H01811 and JPJSCCA20210003.
We acknowledge a financial support by Astrobiology Center.
\end{acknowledgements}


\bibliography{references}
\bibliographystyle{aasjournalv7}

\end{document}